\documentclass[twocolumn]{aastex62}
\usepackage{mathtools}
\usepackage{amsmath}
\usepackage{subfigure}
\usepackage{color}
\usepackage{fnpct}
\usepackage[multiple]{footmisc}

\received{---}
\revised{---}
\accepted{---}
\submitjournal{ApJS}

\shorttitle{Galaxy clusters}
\shortauthors{Hu Zou et al.}

\begin{document}

\title{Galaxy Clusters from the DESI Legacy Imaging Surveys. I. Cluster Detection}

\correspondingauthor{Hu Zou}
\email{zouhu@nao.cas.cn}
\author[0000-0002-6684-3997]{Hu Zou}
\affil{Key Laboratory of Optical Astronomy, National Astronomical Observatories, Chinese Academy of Sciences, Beijing 100012, China}
\author{Jinghua Gao}
\affil{Key Laboratory of Optical Astronomy, National Astronomical Observatories, Chinese Academy of Sciences, Beijing 100012, China}
\author{Xin Xu}
\affil{Key Laboratory of Optical Astronomy, National Astronomical Observatories, Chinese Academy of Sciences, Beijing 100012, China}
\affil{School of Astronomy and Space Science, University of Chinese Academy of Sciences, Beijing 101408, China}
\author{Xu Zhou}
\affil{Key Laboratory of Optical Astronomy, National Astronomical Observatories, Chinese Academy of Sciences, Beijing 100012, China}
\author{Jun Ma}
\affil{Key Laboratory of Optical Astronomy, National Astronomical Observatories, Chinese Academy of Sciences, Beijing 100012, China}
\affil{School of Astronomy and Space Science, University of Chinese Academy of Sciences, Beijing 101408, China}
\author{Zhimin Zhou}
\affil{Key Laboratory of Optical Astronomy, National Astronomical Observatories, Chinese Academy of Sciences, Beijing 100012, China}
\author{Tianmeng Zhang}
\affil{Key Laboratory of Optical Astronomy, National Astronomical Observatories, Chinese Academy of Sciences, Beijing 100012, China}
\author{Jundan Nie}
\affil{Key Laboratory of Optical Astronomy, National Astronomical Observatories, Chinese Academy of Sciences, Beijing 100012, China}
\author{Jiali Wang}
\affil{Key Laboratory of Optical Astronomy, National Astronomical Observatories, Chinese Academy of Sciences, Beijing 100012, China}
\author{Suijian Xue}
\affil{Key Laboratory of Optical Astronomy, National Astronomical Observatories, Chinese Academy of Sciences, Beijing 100012, China}



\begin{abstract}
Based on the photometric redshift catalog of \citet{zou19}, we apply a fast clustering algorithm to identify 540,432 galaxy clusters at $z\lesssim1$ in the DESI legacy imaging surveys, which cover a sky area of about 20,000 deg$^2$. Monte-Carlo simulations indicate that the false detection rate of our detecting method is about 3.1\%. The total masses of galaxy clusters are derived using a calibrated richness--mass relation that are based on the observations of X-ray emission and Sunyaev \& Zel'dovich effect. The median redshift and mass of our detected clusters are about 0.53 and $1.23\times10^{14} M_\odot$, respectively. Comparing with previous clusters identified using the data of the Sloan Digital Sky Survey (SDSS), we can recognize most of them, especially those with high richness. Our catalog will be used for further statistical studies on galaxy clusters and environmental effects on the galaxy evolution, etc. 

\end{abstract}



\section{Introduction} 
Galaxy clusters are the most massive gravitationally bound systems in the universe. They are formed on the cosmic web, tracing the large-scale structure. Galaxy clusters are powerful probes for the structure growth and provide important constraints on cosmological parameters \citep{kra12,pla16a}. Galaxy clusters are also excellent astrophysical laboratories for studying the galaxy evolution, galaxy collision, gravitational lense, dark matter, and extreme physics in dense environments \citep{dre80,but84,moo96,lin04,san04,kne11,kra12,ebe14}.

Galaxy clusters are a collection of galaxies that are bonded by the gravity. In addition to stars constituting the luminous matter of galaxies in a cluster, there are multiple observable compositions, including cold gas and dust in galaxies, hot ionized intra-cluster medium (ICM), and dark matter. The multi-component nature of galaxy clusters make them visible across the electromagnetic spectrum. The X-ray emission and Sunyaev \& Zel'dovich effect in microwave band originating from the hot ICM \citep{sun72,voi05,ble15} can be used to detect redshift-independent samples of galaxy clusters \citep{pif11,rei13,pla16b,tar19}. These samples usually present relatively high purity and completeness but tend to be high-mass systems. The most effective detections of galaxy clusters are based on large-scale optical and near-infrared imaging data. A variety of cluster finding techniques have been developed attributing to ongoing and upcoming wide and deep large-scale photometric surveys. Generally, there are two kinds of cluster finding methods: one is to find the overdensity of galaxies in the three-dimensional space \citep{sza11,wen12,gao20} and the other is based on the red-sequence feature \citep{koe07,hao10,ryk14}, which describes the narrow color distribution of red member galaxies in a cluster (also known as the E/S0 redgeline). The red-sequence method utilizes intrinsic color properties of member galaxies, while the overdensity finding method can identify clusters not presenting the red-sequence feature, especially at high redshift. 

In \citet{gao20}, we introduced a new fast clustering algorithm, called as Clustering by Fast Search and Find of Density Peaks (CFSFDP), to identify galaxy clusters from a photometric redshift (photo-z) catalog, which is based on the 7-band photometry of South Galactic Cap U-band Sky Survey \citep[SCUSS;][]{zho16,zou16}, SDSS, and unWISE \citep{lan16} surveys. A total of about 20,000 clusters were discovered over a sky area of 3700 deg$^2$ in the south Galactic cap. The latest data release\footnote{\url{https://www.legacysurvey.org/dr8/}} of the DESI \citep[Dark Energy Spectroscopic Instrument;][]{des16} legacy imaging surveys provide 5-band photometry over a sky area of about 20,000 deg$^2$ in both northern and southern Galactic caps \citep{dey19}. They are composed of three optical components in $grz$ bands and one near-infrared component from the \textit{Wide-field~Infrared~Survey~Explorer} (WISE) satellite. We have obtained a catalog of accurate photo-zs and stellar masses for more than 0.3 billion morphologically classified galaxies with $r < 23$ mag and in the redshift range of $z<1$ \citep{zou19}. Based on this catalog, we intend to identify a large sample of galaxy clusters by applying the CFSFDP algorithm. Combining the existing and future spectroscopic data (e.g. SDSS/eBOSS and DESI), we will further statistically investigate the properties of clusters and their member galaxies, environment effect on galaxy evolution, gas inflow and outflow of ICM, and large-scale structures, etc. 

This paper is the first of our series work and is organized as follows. Section \ref{sec:data} briefly describes the imaging data and photo-z catalog. Section \ref{sec:detection} introduces our detecting method of galaxy clusters. In this section, we apply Monte-Carlo simulations to analyze the false detection rate of our method. The cluster mass and richness are also estimated.  Section \ref{sec:match} presents the cross-matching with the Abell and other cluster catalogs based on SDSS data. Section \ref{sec:summary} gives a summary. Throughout this paper, we assume a $\Lambda$CDM cosmology with $\Omega_m=0.3$, $\Omega_\Lambda=0.7$, and $H_0=70$ km s$^{-1}$ Mpc$^{-1}$.

\section{Imaging data and photo-z} \label{sec:data}
The DESI is planned to conduct a large-scale spectroscopic survey with 5000-fiber robots installed on the focal plane of the 4m Mayall telescope at Kitt Peak, Arizona \citep{des16}. It will measure the redshifts of about 35 million galaxies and quasars over a 5-year period, aiming to explore the structure growth and expansion history of the universe. The DESI legacy imaging surveys are designed to provide spectroscopic targets for DESI. They are compose of three public optical surveys including the Beijing-Arizona Sky Survey \citep{zou17}, the Dark Energy Camera Legacy Survey \citep{blu16}, and the Mayall $z$-band legacy survey \citep{sil16}, which jointly image a sky area of $\sim$14000 deg$^2$ in $g$, $r$, and $z$ bands. The legacy surveys also integrate the latest WISE observations and provide deep force photometry in two WISE W1 and W2 bands \citep{lan16,mei19}. The latest data release (DR8) covers an area of about 20,000 deg$^2$ in both northern and southern Galactic caps, which includes additional data from Dark Energy Survey \citep{desc16}. The optical imaging depths are 1.5--2 mag deeper than the SDSS and the infrared $W1$ and $W2$ bands are about 1--1.5 mag deeper than the official WISE data. 

The DESI legacy imaging surveys provide deep photometry in 5 bands ranging from optical to near-infrared. Based on the 5-band photometry, we have obtained a catalog of accurate photometric redshifts and stellar masses for galaxies classified by the legacy surveys \citep[see more details in][]{zou19}. The catalog is updated to the DESI DR8 release\footnote{\url{http://batc.bao.ac.cn/~zouhu/doku.php?id=projects:desi_photoz:start}}, which contains about 0.3 billion galaxies with $r<23$ mag. The redshift and mass ranges are about $z < 1$ and $8.4<\log(M_*)<11.9$, respectively. The redshift bias is ignorable and the corresponding accuracy is about 0.017. The stellar mass shows internal and external dispersions of about 0.09 and 0.22 dex, respectively. We have applied some quality cuts to get clean photometry when constructing the photo-z catalog. In order to select galaxies with more reliable photo-z estimations, we limit our galaxy samples to meet the following conditions: 
\begin{equation}
    \begin{split}
	  \mathrm{MAG\_ABS\_R} < -20.5, \\
	 \mathrm{PHOTO\_ZERR} < 0.1,
    \end{split}
	  \label{equ-limits}
 \end{equation}
 where MAG\_ABS\_R is the $r$-band absolute magnitude and PHOTO\_ZERR is the photo-z error. The cut of the photo-z error is used to eliminate galaxies with large photo-z uncertainty. The cut of the absolute magnitude is set to keep more luminous galaxies and hence improve the completeness. Finally, a total of 0.18 billion galaxies are left for identifying clusters in the rest of this paper.  

\section{Cluster Identification} \label{sec:detection}
\subsection{Detecting method}
The CFSFDP approach was first proposed by \citet{rod14} and has been widely used for cluster analyses in a wide range of fields. In this approach, the cluster center is the point that having a higher density than their neighbors and meanwhile it is adequately far away from other points having higher densities. It was demonstrated that this approach can intuitively detect clusters and their members and automatically exclude the outliers. The CFSFDP is characterized in the algorithm simplicity and efficiency so that it is very suitable for identifying galaxy clusters in astronomical photometric surveys. We have successfully applied this approach to find clusters in the south Galactic cap with photo-zs based on the multi-wavelength data from SCUSS, SDSS, and unWISE \citep{gao20}. The photometric redshifts are used to project the galaxies onto the 2D sky plane and then the CFSFDP is used to identify the overdensity peaks. We adopt a similar method as used in \citet{gao20}. The details of the cluster detecting process are described as below. 

\begin{enumerate}
\item Galaxies in our photo-z catalog are divided into small equal-area sky patches (or pixels) in the Hierarchical Equal Area and isoLatitude Pixelisation (HEALPix) projection\footnote{\url{http://healpix.jpl.nasa.gov/}}. We adopt the resolution parameter of $\mathrm{nside}=64$, giving each HEALPix pixel area of about 0.89 deg$^2$. The above pixelation is to facilitate the parallelization of cluster detecting.
\item For each galaxy at a given redshift $z$ in a specified HEALPix pixel, we calculate the local density ($\Phi$) and background density ($\Phi_\mathrm{bkg}$). These two quantities are computed from the galaxies in this pixel and its surrounding 8 neighbor pixels, which have a total area of about 8 deg$^2$. The area is large enough to estimate a proper local background. $\Phi$ is computed as the number of galaxies with distance $R <  0.5$ Mpc in a redshift slice of $z\pm0.04(1+z)$. $\Phi_\mathrm{bkg}$ is computed as the number of galaxies with $R > 1$ Mpc (to avoid the possible cluster region) in the same redshift slice. The background density is scaled to the same area as used in the calculation of the local density. Another crucial parameter $D$ in Mpc is defined as the shortest distance of the specified galaxy to other galaxies with higher $\Phi$. 
\item The cluster centers should be located on the overdensity peaks with large enough local density and adequately far away from  other peaks. They are identified as the galaxies with $\Phi > 4\Phi_\mathrm{bkg}$ and $D > 1$ (typical cluster radius is less than 1 Mpc). If there are multiple galaxies with the same local density, we assign the brightest galaxy in $r$ band as the cluster center. This cluster center is not always the brightest cluster galaxy (BCG) that is usually close to the densest region. We simply regard the $r$-band brightest galaxy  within 0.5 Mpc around the density peak as the BCG. Since each galaxy is pointed to the nearest neighbor with higher $\Phi$, the galaxies in the redshift slice can be traced and clustered to the cluster center. The clustered galaxies are roughly considered as the member candidates. With these members, we can calculate the average position and redshift for each cluster.
\item The number of member galaxies within 1 Mpc around the cluster center are calculated, which is subtracted the local background and denoted as $N_\mathrm{1Mpc}$. $N_\mathrm{1Mpc}$ is regarded as the first-order approximate of the cluster richness.  We require that clusters should have $N_\mathrm{1Mpc}>10$. The total $r$-band luminosity of member galaxies around the cluster center ($L_\mathrm{1Mpc}$) is calculated in unit of $L^*$, where $L^*$ is the characteristic luminosity. $L_\mathrm{1Mpc}$ will be used as a proxy of the cluster mass and richness later. When calculating $L_\mathrm{1Mpc}$, we also subtract the background luminosity that is estimated in the same way as $\Phi_\mathrm{bkg}$. Note that the richness and luminosity in this paper are not corrected for the variation of the imaging depth. Such corrections are quite complicated due to the mixture nature of our data from three different optical surveys plus the WISE infrared survey and the selection of the photo-z sample.  Actually, more than 90\% of the area has the $5\sigma$ $r$-band magnitude limit larger than 23.4 mag. As presented in \citet{zou19}, the completeness of galaxies at $r<23$, which is the magnitude cut for the photo-z catalog used in this paper, is higher than 90\%. As a preliminary estimate, the correction of the richness or luminosity would be less than 10\%.
\end{enumerate}

The above process is performed over the photo-z catalog. Finally, we obtain a catalog of 540,432 galaxy clusters at $z \lesssim 1$. The catalog content is described in Appendix \ref{apx:catalog}.  Figure \ref{fig:example} presents four examples of our detected clusters at different redshifts. As can be seen from this figure, the prominent overdensity features of galaxies make them easily identified from our imaging data by using the above cluster finding method. 

\begin{figure*}[!ht]
\centering
\includegraphics[width=0.8\textwidth]{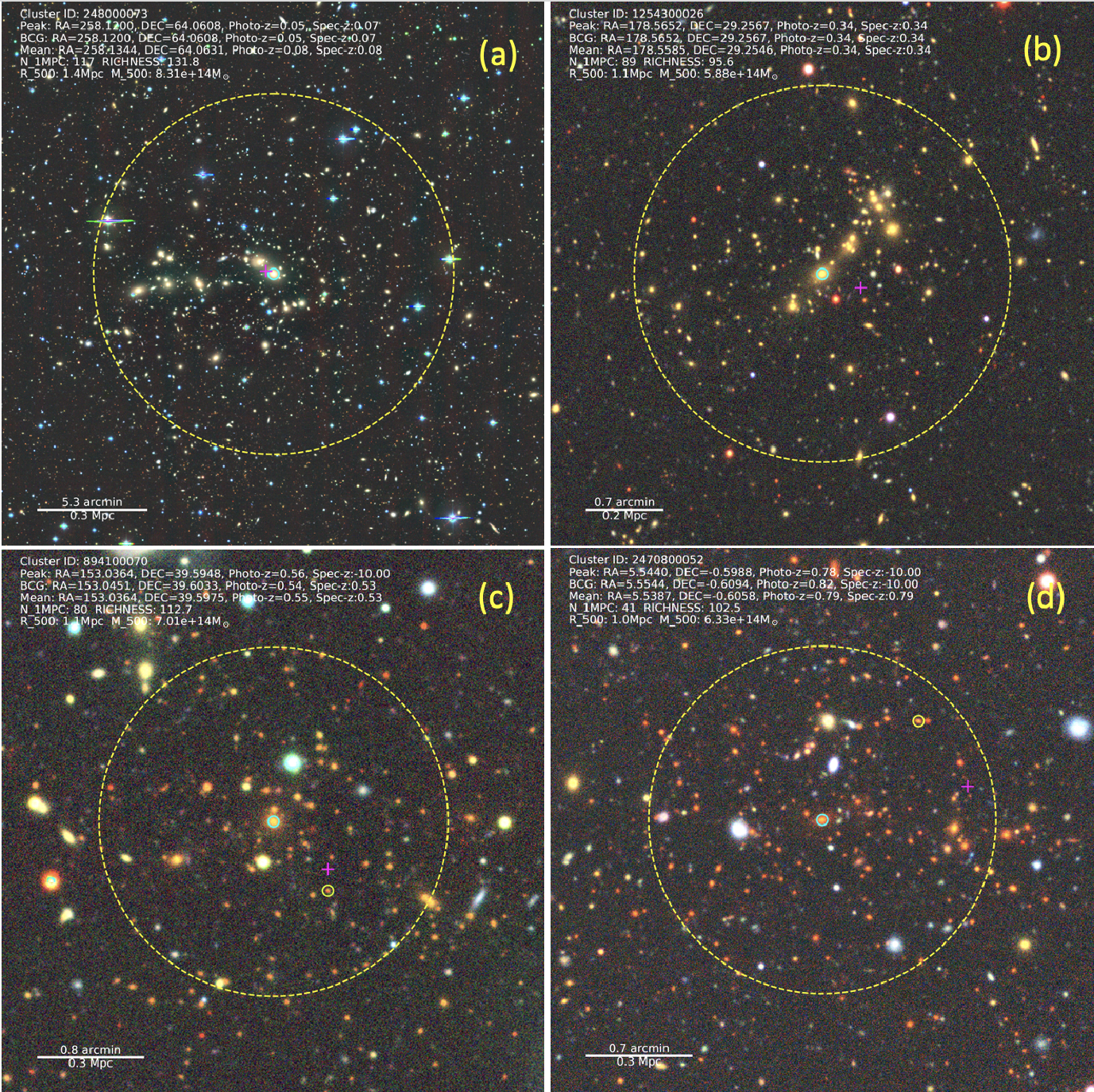}
\caption{Four examples of galaxy clusters at different redshifts: (a) $z=0.05$, (b) $z=0.34$, (c) $z=0.54$, and (d)  $z=0.82$. The small green circle shows the galaxy at the local density peak and the cyan circle at the center marks the BCG. If these two galaxies are the same, the cyan circle will cover the green one. The magenta plus shows the average position of member galaxies. The big yellow dashed circle indicates a radius of 0.5 Mpc. In each panel, we also display the information about the galaxy at the density peak (Peak), the BCG, and average of the members (Mean), including the position (RA, DEC), photo-z (Photo-z), and spectroscopic redshift (Spec-z; if not existing, just giving $-10$). The cluster properties are also presented, including $N_\mathrm{1Mpc}$ (N\_1MPC), richness (RICHNESS), total mass (M\_500), and characteristic radius (R\_500) (see Section \ref{sec:calibration} and Table \ref{tab:cluster}). North is up and east is left.}
\label{fig:example}
\end{figure*}

\subsection{Reliability analysis} \label{sec:falserate}
The cluster detecting process is executed in relatively large redshift slices due to the photo-z uncertainty. It is inevitable to introduce the project effect and hence cause false cluster detections. In addition, the local background fluctuation will also bring contaminations to the low-richness clusters. We perform a Monte-Carlo simulation to evaluate the detecting approach and estimate the false detection rate following a similar method of \citet{wen11}. The simulation is based on the actual photometric data and the specific steps are described as follows. (1) An arbitrary region with an area of 400 deg$^2$ is selected in our photo-z catalog. (2) Galaxies in this region are redistributed on the sky with random walks from original locations in the range of 1--2.5 Mpc and then their redshifts are shuffled. This step can generate new random galaxy backgrounds and meanwhile retain the project effect in some degree.  (3)  The same clustering detecting process used in this paper is applied to these mock galaxies and to finding fictitious clusters. 

\begin{figure*}[!ht]
\centering
\includegraphics[width=\textwidth]{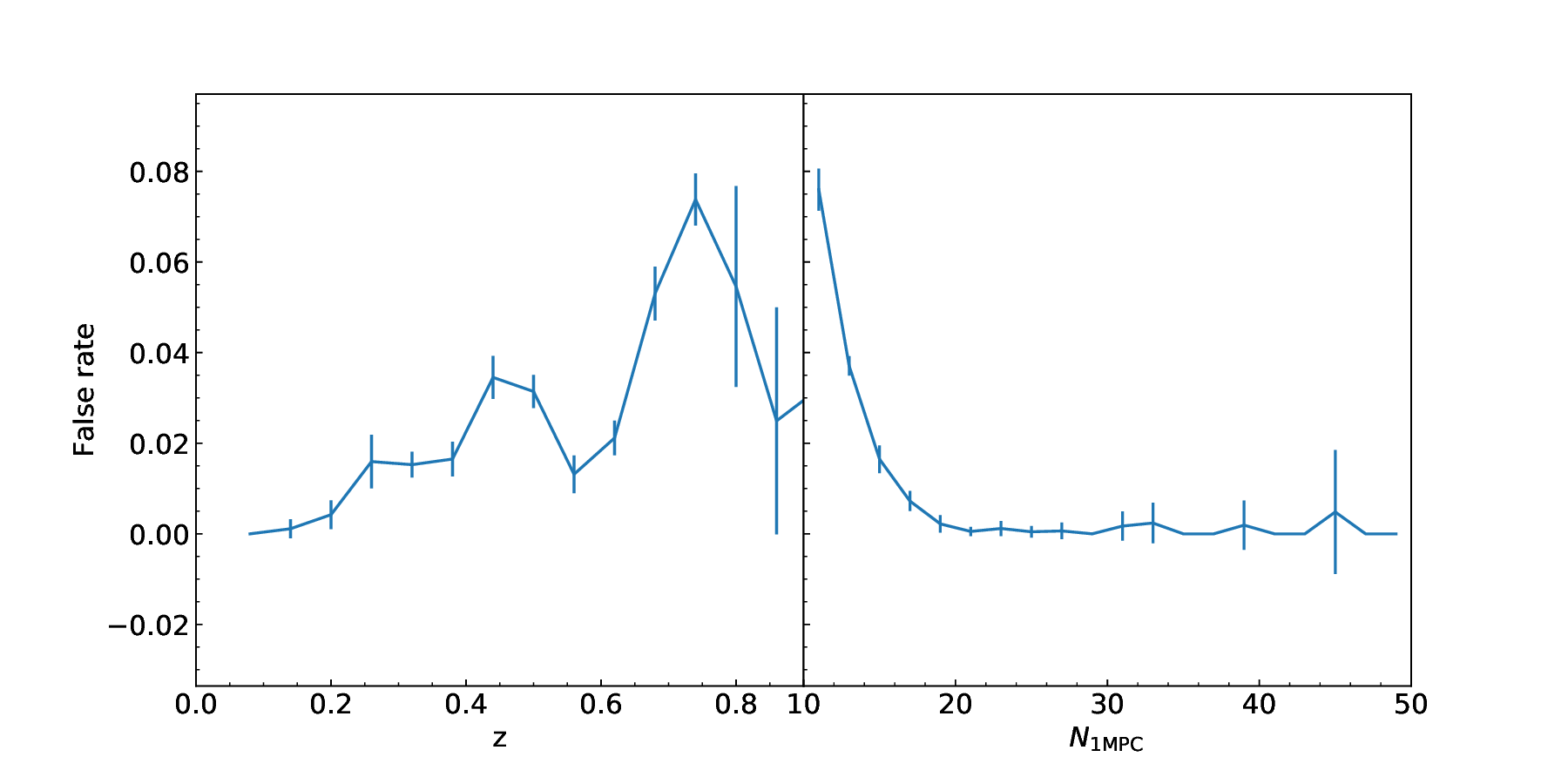}
\caption{Left: false detection rate as function of redshift. Right: false detection rate as function of $N_\mathrm{1Mpc}$. The error bar shows the standard deviation of 10 simulations.}
\label{fig:falserate}
\end{figure*}

We make a total of 10 simulations and identify the false detections in these simulations. The false rate is defined as the ratio of the number of false clusters identified in the simulation to the number of detected clusters in the original catalog. Figure \ref{fig:falserate} presents the average false rate of the above 10 simulations as functions of redshift and $N_\mathrm{1Mpc}$. From this figure, we can see that the false rate generally increases with redshift and decreases with $N_\mathrm{1Mpc}$. In other words, the contamination of our detections is more serious for distant or low-richness clusters. The overall false rate is about 3.1\% and it can go up to about 7-8\% at high redshift and lowest richness.

It should be noticed that the above simulations can estimate the false detections inducing by chance associations and part of false detections caused by the projection effect of uncorrelated structures, because we generate random galaxy backgrounds with limited position changes.  Nevertheless, the projection effects include the impacts from both correlated and uncorrelated large-scale structures.  It is very important to estimate their effect on basic cluster properties (e.g. richness and halo mass) especially when the cluster catalog is used to constrain the cosmological parameters. For example, \citet{cos19}  combined both real data and numerical simulations to characterize the impact of projection effects on the redMaPPer cluster catalog and they found that the projection effect obviously biases the cosmological parameter measurements.  Principally, we can use the mock catalogs from $N$-body simulations to assess the reliability of our cluster finder. However, the synthetic data are model-dependent and are probably biased from realistic observations, such as the halo occupation distribution of cluster galaxies, galaxy color distributions, photometric uncertainty and corresponding photo-z uncertainty. Considering the complexity of the projection effects, we expect an investigation of their impact on the reliability of our detecting method and on the following mass estimation in future work.

\subsection{Mass and richness estimation} \label{sec:calibration}
The X-ray and SZ surveys provide large samples of galaxy clusters at different redshifts, which has reasonably estimates of the cluster mass with relatively small intrinsic scatter. The observables, such as luminosity and temperature in X ray and the integrated Comptonisation parameter (or SZ parameter), are tightly linking to the cluster mass via scaling relations. 

\citet{wen15} collected a sample of 1,192 clusters from the literature and scaled the cluster masses derived by different authors to a unified calibration. These samples are limited to match the galaxy clusters identified from SDSS with the maximum redshift is 0.75. The clusters at high redshift are insufficient, so we supplement the sample of \citet{wen15} with other clusters from X-ray and SZ observations \citep{pif11,tak13,tak14,wen15,tak16,pla16b}.  The cluster masses in different catalogs were estimated using different mass proxies and scaling relations, so it is necessary to recalibrate the masses. So we use the rescaling relations derived by \citet{wen15} to calibrate the cluster mass for these additional data. All the cluster masses are rescaled to the standard of \citet{vik09}. We should note that the published Planck catalog has not applied the Eddington bias correction, which can systematically bias the SZ masses \citep{bat16,med18}. The above recalibration process can reduce this kind of systematic effects. Here, the cluster mass is denoted as $M_{500}$, which is the mass within a characteristic radius ($R_{500}$). $R_{500}$ is defined as the radius within which the mean density of a cluster is 500 times the critical density of the universe ($\rho_c$). According to the definition, $M_{500}$ and $R_{500}$ satisfy the following relation:
\begin{equation}
M_{500} = \frac{4\pi}{3}R_{500}^3\times500\rho_c. \label{equ:massr}
\end{equation}
We collect 3,157 clusters with reliable estimations of $M_{500}$ and $R_{500}$, which spreads over the whole celestial sphere. They are listed in a separate table of Table \ref{tab:mass} in Appendix \ref{apx:mass}, which we called as the calibration catalog. 

\begin{figure*}[!ht]
\centering 
\includegraphics[width=\textwidth]{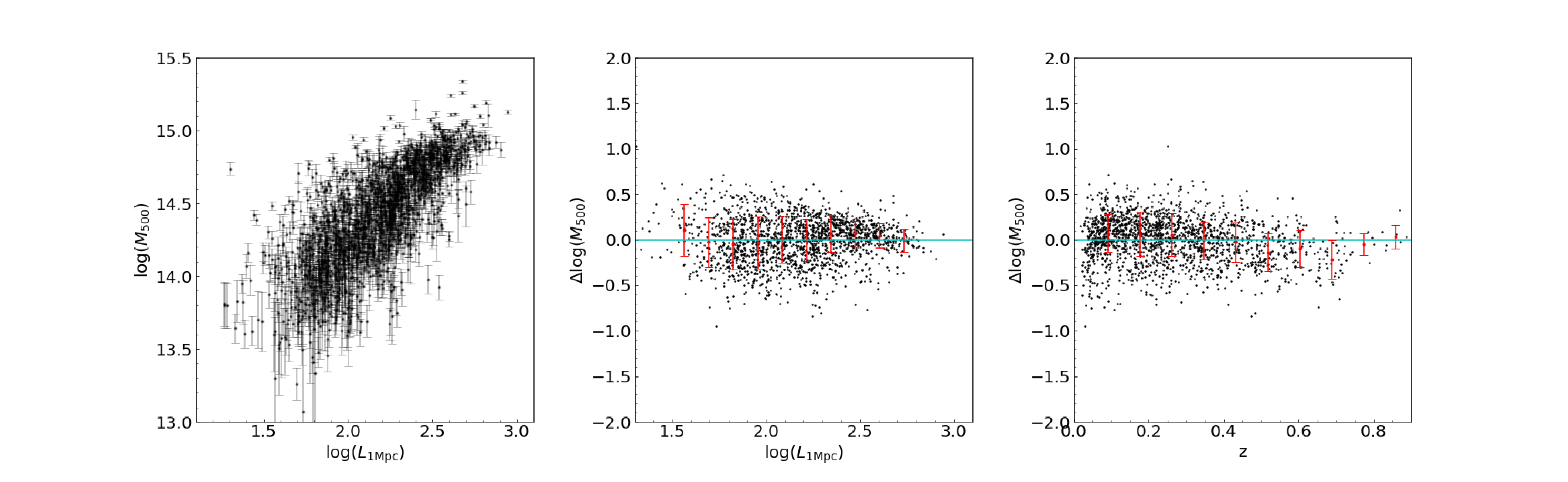}
\caption{Left: correlation between $\log(M_{500}$) and $\log(L_\mathrm{1Mpc})$. The error bar shows the error of $\log(M_{500}$). Middle: the difference ($\Delta \log(M_{500})$) between $\log(M_{500})$ and fitted values  as function of $L_\mathrm{1Mpc}$. The red error bar shows the median and standard deviation in each luminosity bin. Right: $\Delta \log(M_{500})$ as function of redshift. The red error bar shows the median and standard deviation in each redshift bin. The horizontal line displays $\Delta \log(M_{500}) = 0$. }
\label{fig:calibration}
\end{figure*}

As described in \citet{gao20}, $L_\mathrm{1Mpc}$ is used as an optical proxy of the cluster mass. We also use $L_\mathrm{1Mpc}$ to estimate the total mass of our clusters. The galaxy clusters detected in this work are cross-matched with the clusters in the calibration catalog of Table \ref{tab:mass}, following a similar procedure in \citet{ble20}: 1) ranking the clusters in the calibration catalog by the decreasing mass (equivalent to the richness); 2) matching each cluster in the calibration catalog to the richest cluster in our catalog with a redshift error of $\Delta z = 0.06 (1+z)$ and a projected separation error of 1 Mpc; 3) removing each matched cluster from our catalog and repeating the above matching process for all clusters in the calibration catalog.  As a result, we find 1,797 matched clusters that can be used to determine the relation between $L_\mathrm{1Mpc}$ and $M_{500}$. The left panel of Figure \ref{fig:calibration} shows $M_{500}$ as function of $L_\mathrm{1Mpc}$. We can see that there is a tight correlation between these two quantities. The correlation can be described by the following equation:
\begin{equation}
\begin{split}
\log(M_{500}) &= a\log(L_\mathrm{1Mpc})+b\log(1+z) +c, \\
\mathrm{or~}  M_{500} &=10^cL_\mathrm{1Mpc}^a(1+z)^b,
\end{split}\label{equ:calib}
\end{equation}
where $a$, $b$, and $c$ are constants to be determined and the term of $(1+z)$ is regarded as the correction term for the redshift evolution and incompleteness. 

The above luminosity--mass relation suffers from the Malmquist bias as the cluster sample is constructed with flux-limited X-ray and SZ observations. The Malmquist bias makes the sample preferentially contain brighter or massive clusters. It is especially serious for clusters with low richness at high redshift. The compilations of our clusters are from X-ray and SZ observations with different depths, which may mitigate the effect of the Malmquist bias, but also induces the complication of the bias correction. We hope a further investigation of this problem in the future. Considering the bias, we assign the data with weight of $\frac{L_\mathrm{1Mpc}}{z}$ when fitting the data with Equation (\ref{equ:calib}). This will put more weight to the clusters with larger richness and at lower redshift. We get the best-fit parameters via weighted linear regression:  $a=0.81\pm0.02$, $b=0.50\pm0.14$, and $c=12.61\pm0.04$. The middle and right panels of Figure \ref{fig:calibration} presents the residual between observed $M_{500}$ and the fitted one as functions of $L_{500}$ and redshift. The overall RMS of the residual is about 0.2 dex. The residual RMS changes with $L_\mathrm{1Mpc}$, ranging from 0.28 to 0.12 dex. The residual biases at high redshift, partly reflecting the Malmquist bias. We apply the Equation (\ref{equ:calib}) to estimate $M_{500}$ and  Equation (\ref{equ:massr}) to estimate $R_{500}$ for our detected clusters. We also use the Equation (17) in \citet{wen15} to estimate the richness. These parameters are included in Table \ref{tab:cluster}.  

\subsection{Cluster statistics and photo-z accuracy}
Figure \ref{fig:stats} shows the parameter distributions for the whole 540,432 galaxy clusters detected in this paper.  The median redshift is about 0.53 and the number density decreases at $z> \sim0.6$, where the cluster detection should be more incomplete. The median richness is about 22.5. The range of the total mass $\log(M_{500})$ is 13.5--14.8 and the median is about 14.1 (equivalent to $1.23\times10^{14} M_\odot$). The characteristic radius of $R_{500}$ ranges from about 0.4 Mpc to 1.0 Mpc and the median value is about 0.63 Mpc. 
\begin{figure}[!ht]
\centering 
\includegraphics[width=\columnwidth]{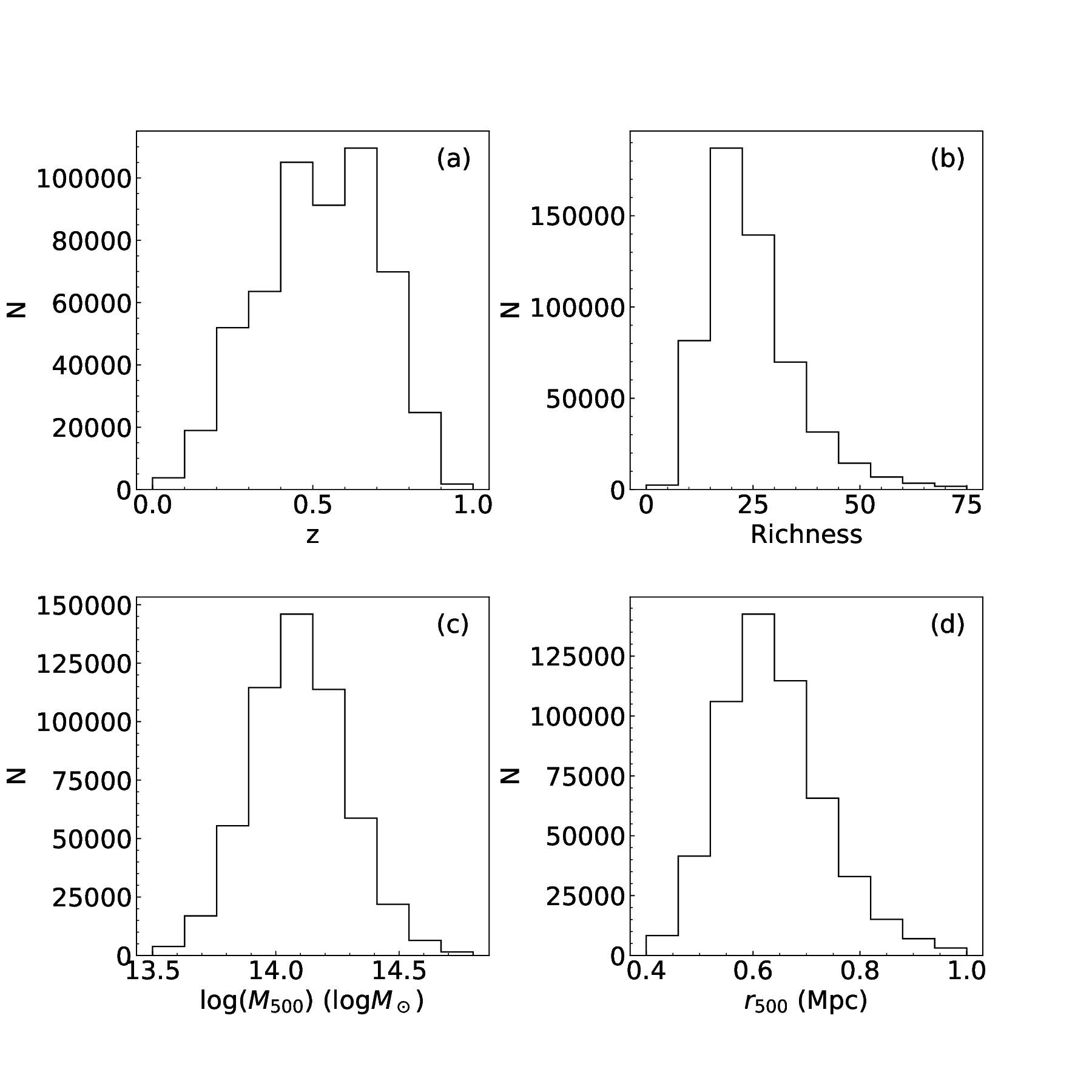}
\caption{Distributions of redshift (a), richness (b), $\log(M_{500})$ (c), and $R_{500}$ (d) for our clusters.}
\label{fig:stats}
\end{figure}

\begin{figure}[thb]
\centering
\includegraphics[width=\columnwidth]{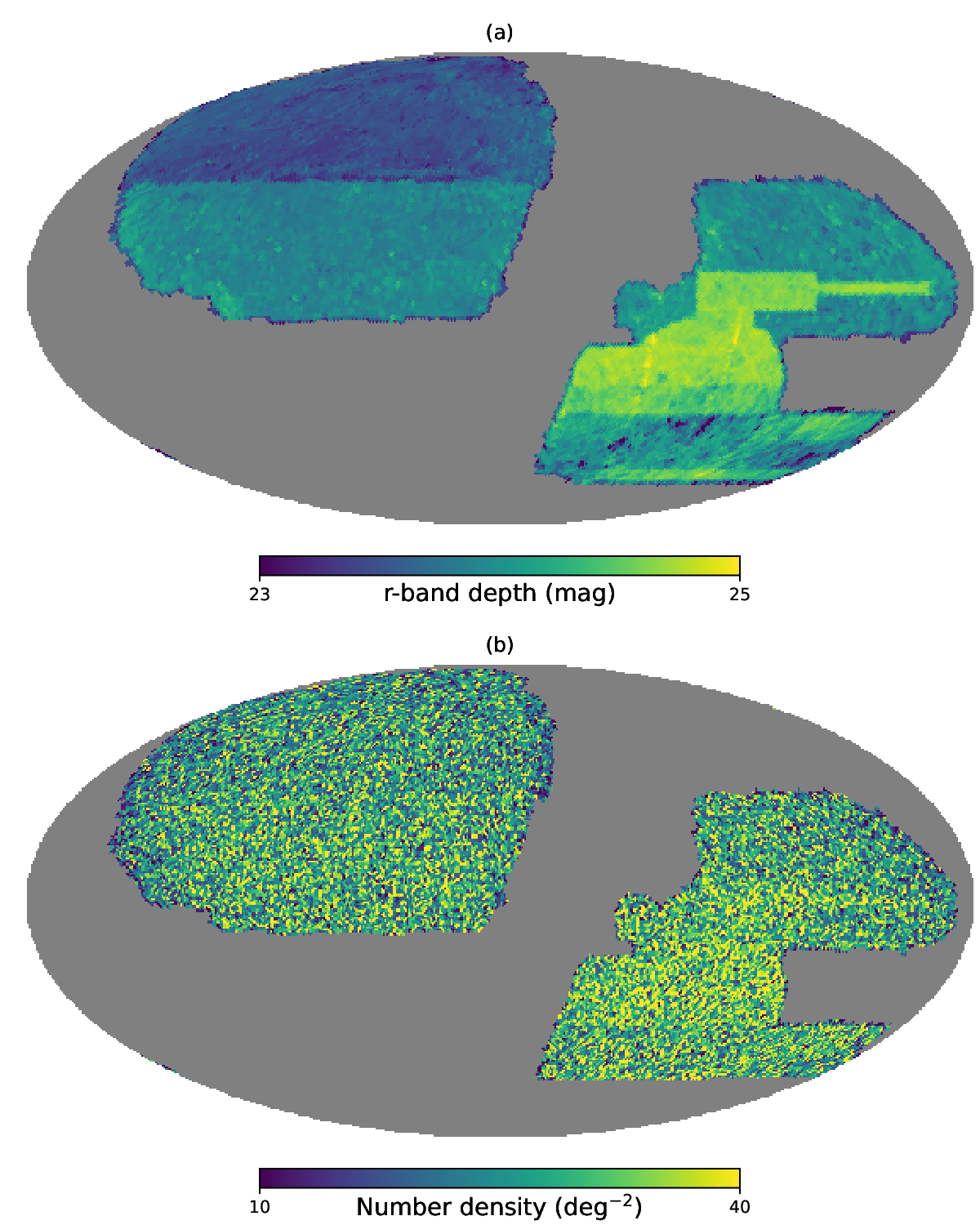}
\caption{(a) The $r$-band depth map in Mollview projection. (b) Number density distribution of our detected clusters.}
\label{fig:coverage}
\end{figure}

\begin{figure}[thb]
\centering
\includegraphics[width=\columnwidth]{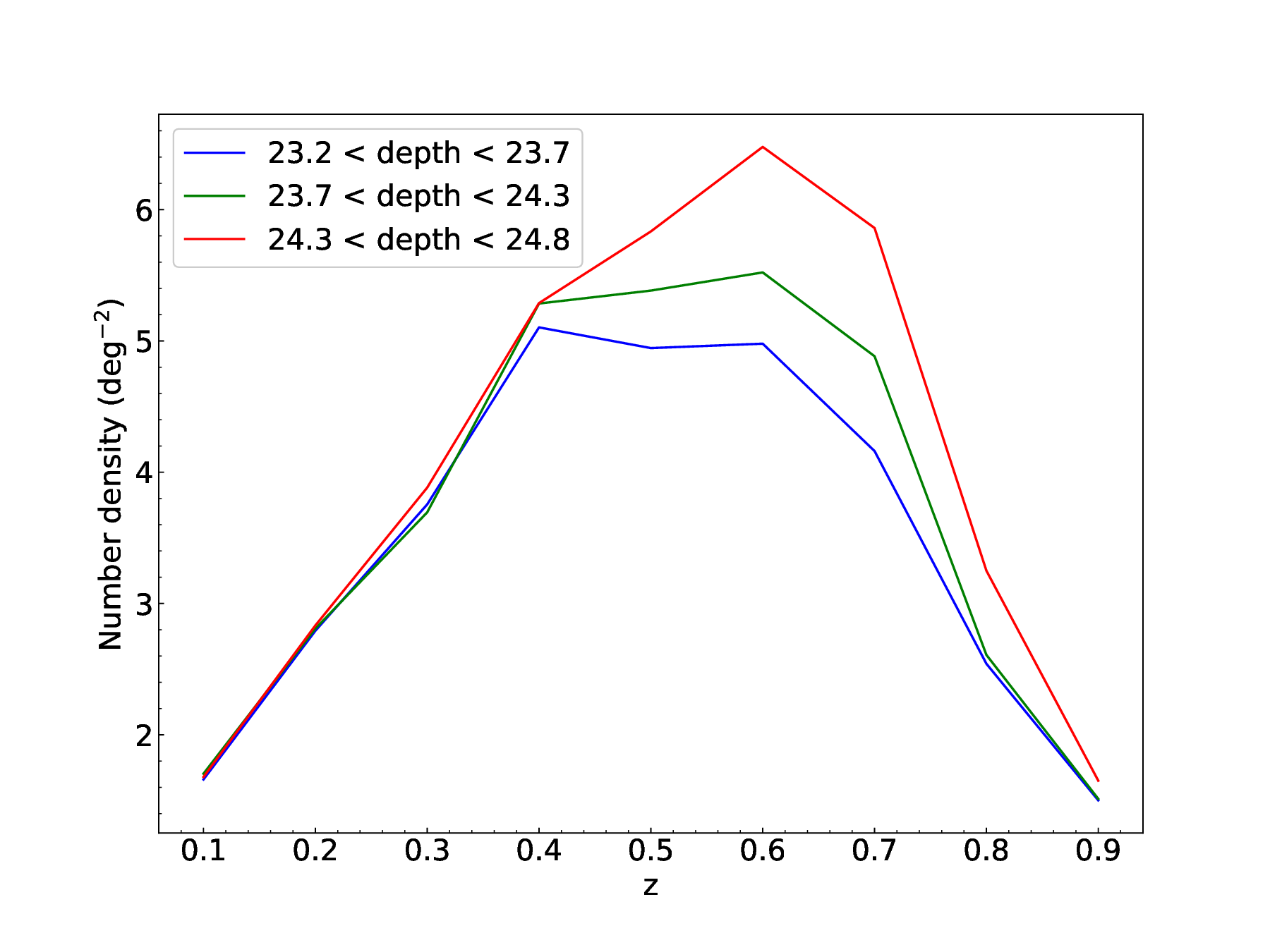}
\caption{Number density as function of redshift for three subsamples of our clusters under different $r$-band imaging depths. The redshift bin is 0.1.}
\label{fig:numden}
\end{figure}

Figure \ref{fig:coverage} displays both the distributions of $r$-band depth and number density of our clusters over the DESI imaging footprint.  Note that the depth is referred to $5\sigma$ $r$-band magnitude limit for extended sources with the correction of the Galactic extinction. The depth map in Figure \ref{fig:coverage}a shows three distinct regions with different imaging depths, which reflects that the observing data were obtained by different facilities. The general depth differences among these three components are about 0.6 mag.  Although the depth is inhomogeneous across the survey footprint, we still obtain uniformly distributed cluster samples attributing to our relatively conservative magnitude cut of our photo-z catalog (see Figure \ref{fig:coverage}b). In order to more clearly show the effect of the imaging depth on our cluster detection, we present the number densities as function of redshift for three subsamples of our clusters under different $r$-band depths in Figure \ref{fig:numden}. From this figure, we can see that the imaging depth only affects the cluster detection at $z>0.4$. More clusters are detected in the deeper area and it becomes more pronounced as the redshift increases.

\begin{figure}[!ht]
\centering 
\includegraphics[width=\columnwidth]{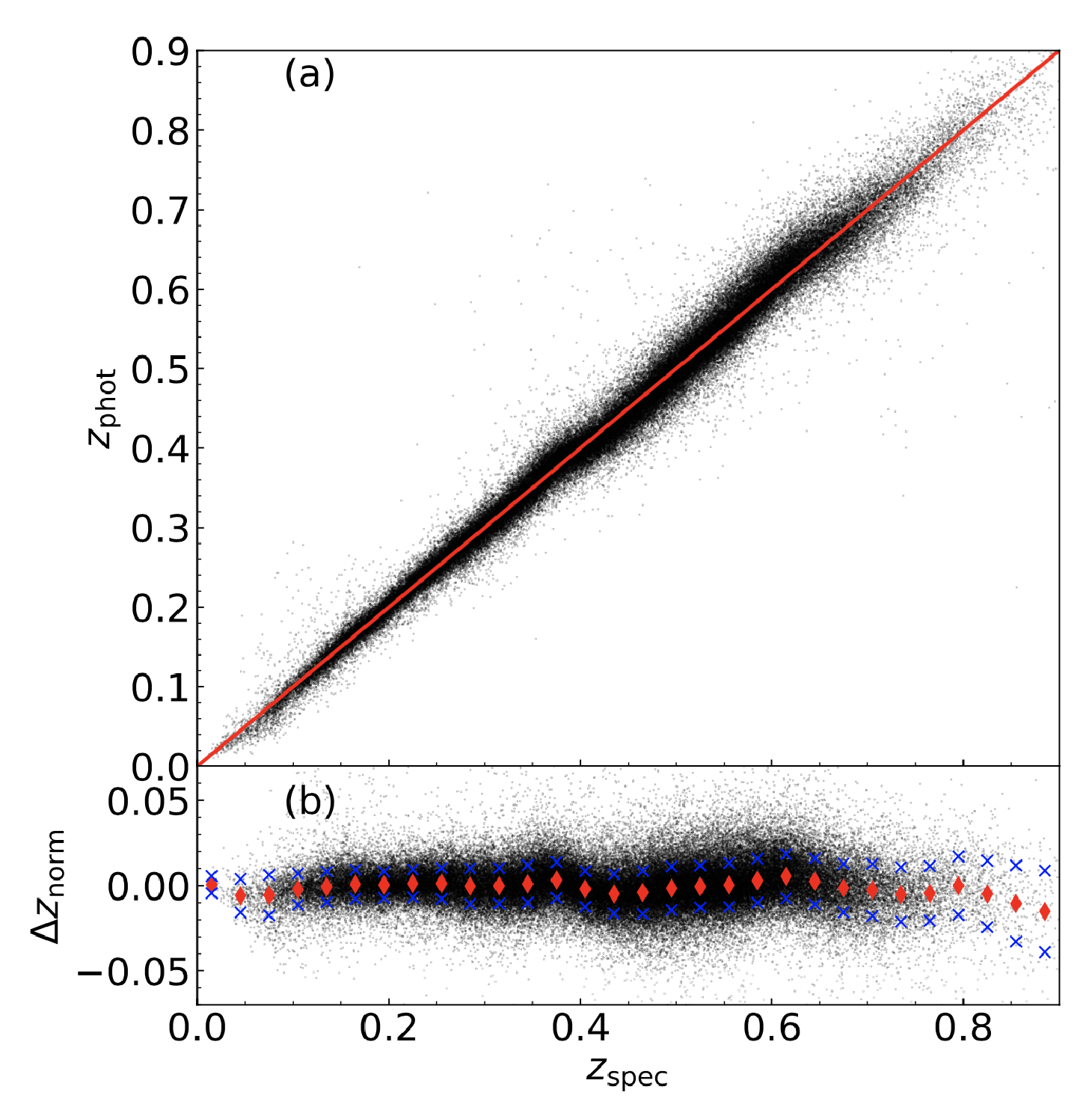}
\caption{(a) Comparison between photometric and spectroscopic redshifts of the BCGs in our cluster catalog. The diagonal line denotes $z_\mathrm{phot}=z_\mathrm{spec}$. (b)  $\sigma_{\Delta z_\mathrm{norm}}$ as function of $z_\mathrm{spec}$. The red diamonds and blue crosses display the medians and standard deviations in different redshift bins.\label{fig:zcomp}}
\label{fig:stats}
\end{figure}

\begin{figure}[!ht]
\centering 
\includegraphics[width=\columnwidth]{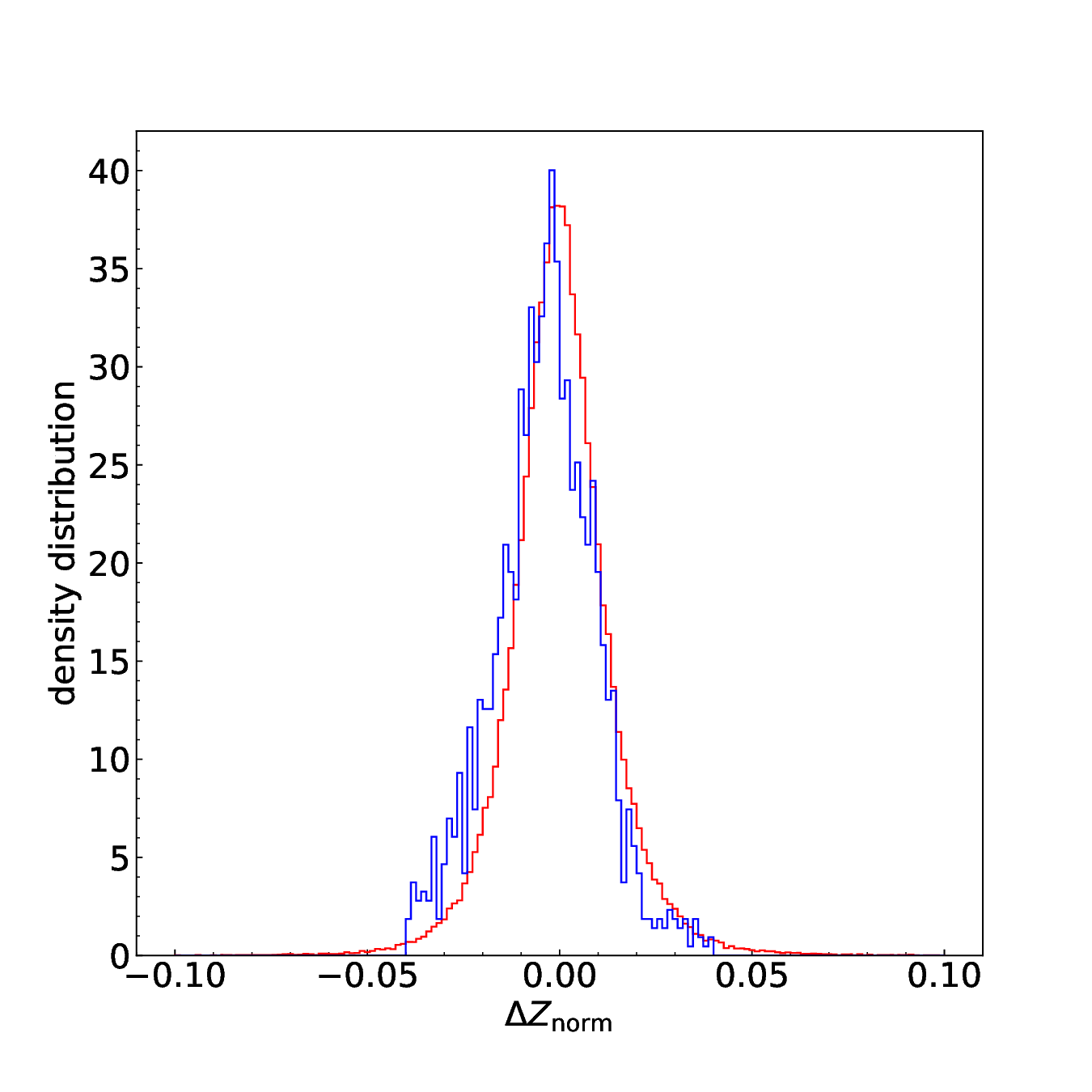}
\caption{The $\sigma_{\Delta z_\mathrm{norm}}$ distribution (red) of the BCGs in our clusters, which have spectroscopic redshift measurements. The blue histogram is the $\sigma_{\Delta z_\mathrm{norm}}$ for confirmed galaxy clusters with reliable redshift measurements, which are detected in X-ray and SZ-effect observations.\label{fig:zhist}}
\label{fig:stats}
\end{figure}

The photo-z accuracy of our galaxy clusters are estimated by comparing the photometric redshifts of the BCGs in our catalog with spectroscopic redshifts collected by \citet{zou19}. There are a total of 12,2390 BCGs with spectroscopic redshifts. The top panel of Figure \ref{fig:zcomp} shows the comparison between the photometric redshift ($z_\mathrm{phot}$) and spectroscopic redshift ($z_\mathrm{spec}$). The bottom panel presents the photo-z accuracy of $\Delta z_\mathrm{norm} = (z_\mathrm{phot}-z_\mathrm{spec})/(1+z_\mathrm{spec})$ as function of $z_\mathrm{spec}$. We can see that the photo-z accuracy changes from about 0.01 at $z\sim0.1$ to about 0.024 at $z\sim 0.9$. The redshift histogram in Figure \ref{fig:zhist} shows the distribution of $\sigma_{\Delta z_\mathrm{norm}}$ of the total sample. The overall photo-z accuracy is $\sigma_{\Delta z_\mathrm{norm}} = 0.012$. In addition, most of the galaxy clusters used for mass calibration as described in Section \ref{sec:calibration} have reliable redshifts. We also estimate our cluster photo-z accuracy using these confirmed clusters, which is shown in the blue histogram of Figure \ref{fig:zhist}. Such a comparison also gives a similar photo-z accuracy of $\sigma_{\Delta z_\mathrm{norm}} = 0.013$. 

\section{Matching with other cluster catalogs} \label{sec:match}
The most famous cluster catalog is the Abell catalog, which was constructed by the visual identification on the photographic plates. There are also several catalogs of galaxy clusters identified from the SDSS photometric data. These catalogs are either based on the red-sequence feature or the overdensity feature in the photo-z space. Through matching our catalog with these catalogs, we can check the detecting repeatability of the clusters from both shallower photometric data and deeper DESI imaging data. Such comparisons can also indicate the reliability of our cluster detection. The Abell clusters were visually identified as galaxy overdensities  \citep{abe89}. Among other catalogs that are based on SDSS data, the clusters of MaxBCG \citep{koe07}, GMBCG \citep{hao10},  and redMaPPer \citep{ryk14} were identified through the red-sequence feature, while the clusters of AMF \citep{sza11}  and WHL15 \citep{wen15} were obtained using the overdensity feature. We summarize the redshift range, adopted method and number of clusters in the DESI footprint of these catalogs in Table \ref{tab:matching}. If the clusters in the Abell catalog have spectroscopic redshifts, we use their spectroscopic redshifts as the distance indicator. For those clusters without redshifts in the Abell catalog, we only use our photo-z as the distance tracer to match the clusters.  The projected separation is limited to 1 Mpc, which is set to be a little larger than the characteristic radius ($R_{500}$ of most clusters as shown in Figure \ref{fig:stats}).  For the SDSS cluster catalogs, we match them with our cluster catalog using a redshift tolerance of $\Delta z<0.06 (1+z)$ and the same separation tolerance. The larger the separation tolerance is set, the more real clusters would be matched in consideration of the uncertainty of the cluster center, while more possible false association would occur and vice versa. The matching results with our catalog are listed in Table \ref{tab:matching}.

\begin{table*}[!ht]
\begin{center}
\caption{Matching our catalog with other cluster catalogs} \label{tab:matching}
 \begin{tabular}{cccccc}
 \tableline \tableline
 Catalog & Redshift & Method & Number & Matched  & Percent  \\
 (1) & (2) &   (3)  & (4)  & (5)  & (6)   \\  \\   
 \tableline
   Abell & \nodata & visual overdensity & 2,298 & 2,143 & 93.3\% \\
   MaxBCG &$0.1<z<0.3$ & red sequence &  13,806 & 10,494 & 76.0\% \\
   GMBCG &$0.1<z<0.55$ & red sequence &   55,317 & 29,980 & 54.2\% \\
   AMF &$0.045<z<0.78$ & overdensity  & 44,692 & 33,731 & 75.5\%  \\
   redMaPPer &$0.08<z<0.55$ & red sequence & 25,835 & 24,645, & 95.4\%  \\
   WHL15 &$0.05<z<0.8$ & overdensity & 121,480 & 80,686 & 66.4\% \\
  \tableline
\end{tabular}
\end{center}
\tablecomments{(1) Cluster catalogs:  Abell \citep{abe89}, MaxBCG \citep{koe07}, GMBCG \citep{hao10}, AMF \citep{ban18}, redMaPPer \citep{ryk14}, and WHL15 \citep{wen15}. (2) Redshift range. (3) Adopted method for identifying clusters. (4) Number of clusters in the DESI imaging footprint. (5) Number of matched clusters using a matching separation tolerance of 2 Mpc. (6) Matching rate using a matching separation tolerance of 2 Mpc. (7) Number of matched clusters using a matching separation tolerance of 1 Mpc. (8) Matching rate using a matching separation tolerance of 1 Mpc. }
\end{table*}

\begin{figure*}[!ht]
\centering 
\includegraphics[width=0.8\textwidth]{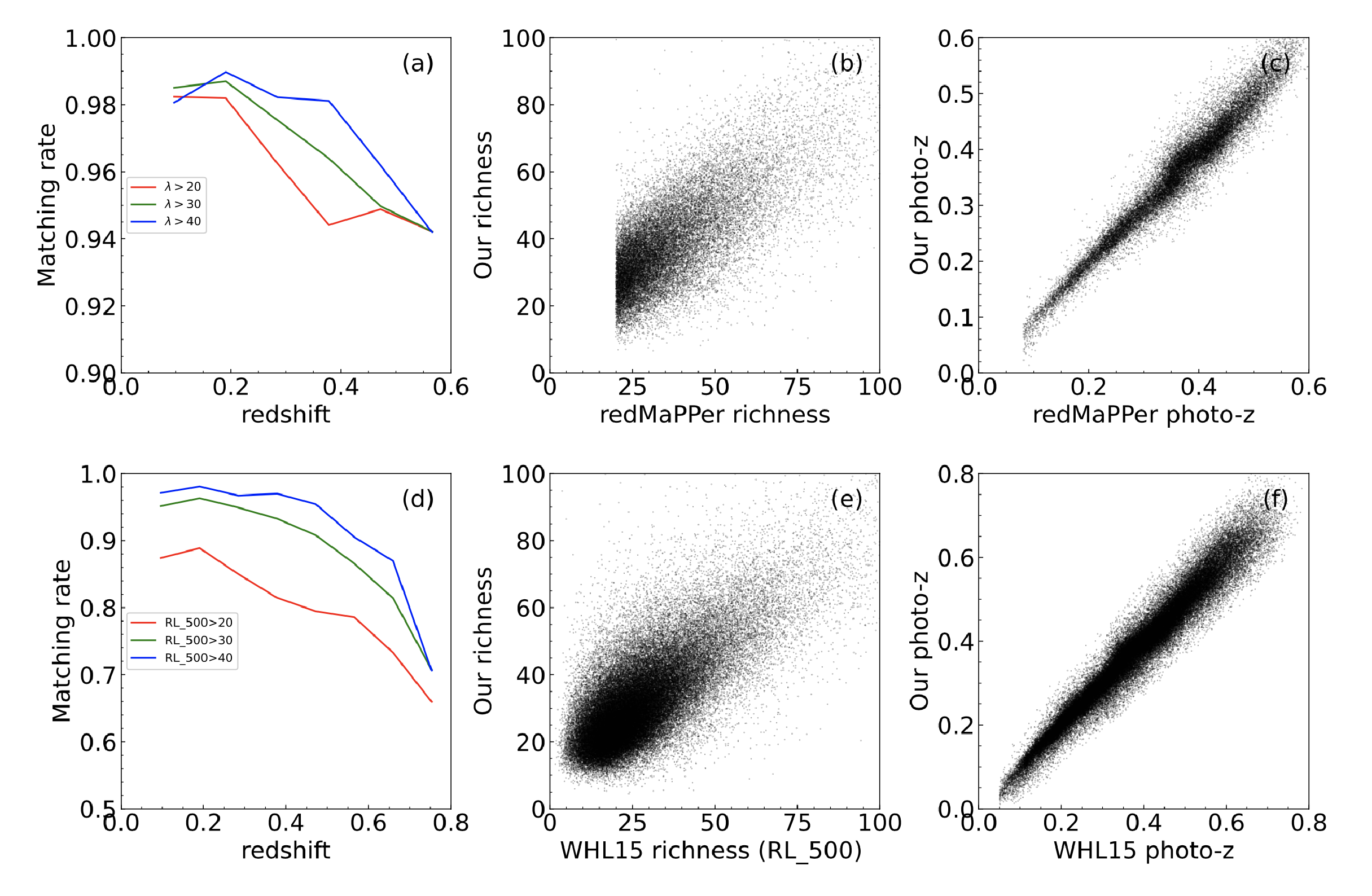}
\caption{(a) The matching rate between the redMaPPer catalog and our catalog as function of redshift and richness. (b) The correlation between the richness of our clusters and that of the redMaPPer clusters ($\lambda)$. (c) The correlation between the photo-z of our clusters and that of the redMaPPer clusters. (d)  The matching rate between the WHL15 catalog and our catalog as function of redshift and richness. (e) The correlation between the richness of our clusters and that of the WHL15 clusters (RL\_500). (f) The correlation between the photo-z of our clusters and that of the WHL15 clusters. \label{fig:catcomp}}
\label{fig:stats}
\end{figure*}

As checked, most of the mismatching clusters in those catalogs based on the SDSS data are the clusters with lower richness or at higher redshift, which are possible false detections. We select two representative catalogs, redMaPPer and WHL15, to show the matching rates as function of redshift and richness and present the comparison of richness and redshift with those in our catalog in Figure \ref{fig:catcomp}. Generally, the matching rate increases as the richness increases or the redshift decreases. The richnesses among these catalogs show good correlations, although there are somewhat large scatters. The photo-z also shows pretty correlations. Compared to the WHL15 photo-z, the photo-z of the redMaPPer clusters is more consistent with that of our clusters.


\section{Summary} \label{sec:lens} \label{sec:summary}
Galaxy clusters are excellent probes for studying galaxy formation and evolution in the dense environments and they can be also used to constrain cosmological parameters. Using the photometric redshift catalog of \citet{zou19}, we identify a catalog of galaxy clusters over the 20000-deg$^2$ sky footprint of the DESI legacy imaging surveys. This sample of clusters will be used to statistically study the galaxy clusters at $z < 1$, the galaxy environment, ICM, large-scale structures of the universe, etc. As the first of a series work, this paper is focused on the cluster detection. 

In total, we identify 540,432 galaxy clusters with $z\lesssim1$. These clusters are uniformly distributed over the imaging footprint. Monte-Carlo simulations show that the false detection rate is about 3.1\% and it becomes worse as the redshift increases or the richness decreases. The false rate at high redshift or low richness can reach up to about 8\%. We utilize the mass measurements from X-ray and radio observations to calibrate the total mass and richness of the detected clusters by using the optical luminosity $L_\mathrm{1Mpc}$. The median mass and richness are about $1.23\times10^{14} M_\odot$ and 22.5, respectively.  Comparing our catalog with the Abell catalog, we can recover almost all Abell clusters. We also compare our detected clusters with those based on the SDSS data that are about 2 magnitude shallower than the DESI imaging surveys. It is found that we can recognize most of the clusters identified from SDSS, especially those with high richness. 

\acknowledgments

We thank the anonymous referee for his/her thoughtful comments and insightful suggestions that improve our paper greatly. This work is supported by Major Program of National Natural Science Foundation of China (No. 11890691, 11890693). It is also supported by the National Natural Science Foundation of China (NSFC; grant Nos. 11733007, 11673027, 11873053, 12073035) and the National Key R\&D Program of China No. 2019YFA0405501.

The Legacy Surveys consist of three individual and complementary projects: the Dark Energy Camera Legacy Survey (DECaLS; NOAO Proposal ID \# 2014B-0404; PIs: David Schlegel and Arjun Dey), the Beijing-Arizona Sky Survey (BASS; NOAO Proposal ID \# 2015A-0801; PIs: Zhou Xu and Xiaohui Fan), and the Mayall z-band Legacy Survey (MzLS; NOAO Proposal ID \# 2016A-0453; PI: Arjun Dey). DECaLS, BASS and MzLS together include data obtained, respectively, at the Blanco telescope, Cerro Tololo Inter-American Observatory, National Optical Astronomy Observatory (NOAO); the Bok telescope, Steward Observatory, University of Arizona; and the Mayall telescope, Kitt Peak National Observatory, NOAO. The Legacy Surveys project is honored to be permitted to conduct astronomical research on Iolkam Du'ag (Kitt Peak), a mountain with particular significance to the Tohono O'odham Nation.

NOAO is operated by the Association of Universities for Research in Astronomy (AURA) under a cooperative agreement with the National Science Foundation.

This project used data obtained with the Dark Energy Camera (DECam), which was constructed by the Dark Energy Survey (DES) collaboration. Funding for the DES Projects has been provided by the U.S. Department of Energy, the U.S. National Science Foundation, the Ministry of Science and Education of Spain, the Science and Technology Facilities Council of the United Kingdom, the Higher Education Funding Council for England, the National Center for Supercomputing Applications at the University of Illinois at Urbana-Champaign, the Kavli Institute of Cosmological Physics at the University of Chicago, Center for Cosmology and Astro-Particle Physics at the Ohio State University, the Mitchell Institute for Fundamental Physics and Astronomy at Texas A\&M University, Financiadora de Estudos e Projetos, Fundacao Carlos Chagas Filho de Amparo, Financiadora de Estudos e Projetos, Fundacao Carlos Chagas Filho de Amparo a Pesquisa do Estado do Rio de Janeiro, Conselho Nacional de Desenvolvimento Cientifico e Tecnologico and the Ministerio da Ciencia, Tecnologia e Inovacao, the Deutsche Forschungsgemeinschaft and the Collaborating Institutions in the Dark Energy Survey. The Collaborating Institutions are Argonne National Laboratory, the University of California at Santa Cruz, the University of Cambridge, Centro de Investigaciones Energeticas, Medioambientales y Tecnologicas-Madrid, the University of Chicago, University College London, the DES-Brazil Consortium, the University of Edinburgh, the Eidgenossische Technische Hochschule (ETH) Zurich, Fermi National Accelerator Laboratory, the University of Illinois at Urbana-Champaign, the Institut de Ciencies de l'Espai (IEEC/CSIC), the Institut de Fisica d'Altes Energies, Lawrence Berkeley National Laboratory, the Ludwig-Maximilians Universitat Munchen and the associated Excellence Cluster Universe, the University of Michigan, the National Optical Astronomy Observatory, the University of Nottingham, the Ohio State University, the University of Pennsylvania, the University of Portsmouth, SLAC National Accelerator Laboratory, Stanford University, the University of Sussex, and Texas A\&M University.

BASS is a key project of the Telescope Access Program (TAP), which has been funded by the National Astronomical Observatories of China, the Chinese Academy of Sciences (the Strategic Priority Research Program "The Emergence of Cosmological Structures" Grant \# XDB09000000), and the Special Fund for Astronomy from the Ministry of Finance. The BASS is also supported by the External Cooperation Program of Chinese Academy of Sciences (Grant \# 114A11KYSB20160057), and Chinese National Natural Science Foundation (Grant \# 11433005).

The Legacy Survey team makes use of data products from the Near-Earth Object Wide-field Infrared Survey Explorer (NEOWISE), which is a project of the Jet Propulsion Laboratory/California Institute of Technology. NEOWISE is funded by the National Aeronautics and Space Administration.

The Legacy Surveys imaging of the DESI footprint is supported by the Director, Office of Science, Office of High Energy Physics of the U.S. Department of Energy under Contract No. DE-AC02-05CH1123, by the National Energy Research Scientific Computing Center, a DOE Office of Science User Facility under the same contract; and by the U.S. National Science Foundation, Division of Astronomical Sciences under Contract No. AST-0950945 to NOAO.

\appendix

\section{Information about our cluster catalog} \label{apx:catalog}
In total, we obtain 540,432 galaxy cluster with a false detection rate of about 3.1\%. Table \ref{tab:cluster} lists the content contained in our catalog. It includes the position and redshift of the cluster center, which is the galaxy at the local density peak. The catalog also includes the basic information of the BCG including the position, redshift, observed magnitudes, absolute magnitudes, and stellar mass, which are inherited from the photo-z catalog\footnote{\url{http://batc.bao.ac.cn/~zouhu/doku.php?id=projects:desi_photoz:start}}.  The cluster properties estimated in this paper are also contained in this catalog.  This catalog will be uploaded to CDS and it is also available at \url{http://batc.bao.ac.cn/~zouhu/doku.php?id=projects:desi_clusters:start}.
\begin{table}[!ht]
\begin{center}
\tiny
\caption{Column description for our cluster catalog} \label{tab:cluster}
 \begin{tabular}{lccl}
 \tableline
 \tableline
 Column & Unit & Format  & Description \\
  \tableline
 CLUSTER\_ID &  & Long & Cluster ID \\
 RA\_PEAK & degree & Double &  R.A. for the density peak (J2000) \\
 DEC\_PEAK & degree & Double & decl. for the density peak (J2000) \\
 PZ\_PEAK &  &  Float & Photometric redshift for the density peak \\
 SZ\_PEAK &  & Float & Spectroscopic redshift for the density peak if existing \\
 DEN\_PEAK & & Integer & Local density $\Phi$ for the density  peak \\
 BKG\_PEAK & & Float  & Local background density $\Phi_\mathrm{bkg}$ for the density peak \\
 RA\_MEAN & degree & Double & Mean R. A. of possible members \\
 DEC\_MEAN & degree & Double & Mean decl. of possible members \\ 
 N\_1MPC &  & Integer & Number of member galaxies within 1 Mpc from the cluster center \\
 L\_1MPC & $L^*$  & Float & Total luminosity of member galaxies within 1 Mpc from the cluster center \\
 M\_500 & $\log_{10}(M_\odot)$ & Float & Total mass of the cluster $M_{500}$ \\
 R\_500 & Mpc & Float & Characteristic radius $R_{500}$ \\
 RICHNESS & & Float &  Cluster richness \\
 RA\_BCG & degree & Double & R.A. for the BCG (J2000) \\
 DEC\_BCG & degree & Double & decl. for the BCG (J2000) \\
 PZ\_BCG & & Float & Photometric redshift for the BCG \\
 PZERR\_BCG &  & Float & Photometric redshift error for the BCG \\ 
 SZ\_BCG &  & Float & Spectroscopic redshift for the BCG if existing \\ 
 MAG\_G\_BCG & mag & Float & $g$-band magnitude for the BCG \\
 MAG\_R\_BCG &  mag & Float & $r$-band magnitude for the BCG \\
 MAG\_Z\_BCG &  mag & Float & $z$-band magnitude for the BCG \\
 MAG\_W1\_BCG &  mag & Float & $W1$-band magnitude for the BCG \\
 MAG\_W2\_BCG &  mag & Float & $W2$-band magnitude for the BCG \\
 MAGERR\_G\_BCG &  mag & Float & $g$-band magnitude error for the BCG \\
 MAGERR\_R\_BCG &  mag & Float & $r$-band magnitude error for the BCG \\ 
 MAGERR\_Z\_BCG &  mag & Float & $z$-band magnitude error for the BCG \\ 
 MAGERR\_W1\_BCG &  mag & Float & $W1$-band magnitude error for the BCG \\
 MAGERR\_W2\_BCG &  mag & Float & $W2$-band magnitude error for the BCG \\ 
 MAG\_ABS\_G\_BCG &  mag & Float & $g$-band absolute magnitude for the BCG \\ 
 MAG\_ABS\_R\_BCG &  mag & Float & $r$-band absolute magnitude for the BCG \\ 
 MAG\_ABS\_Z\_BCG &  mag & Float & $z$-band absolute magnitude for the BCG \\ 
 MAG\_ABS\_W1\_BCG &  mag & Float & $W1$-band absolute magnitude for the BCG \\
 MAG\_ABS\_W2\_BCG &  mag & Float & $W2$-band absolute magnitude for the BCG \\ 
 MASS\_BEST\_BCG & $\log_{10}(M_\odot)$ & Float & Logarithmic stellar mass for the BCG \\ 
 MASS\_INF\_BCG & $\log_{10}(M_\odot)$  & Float & Lower limit of logarithmic stellar mass with 68\% confidence level for the BCG \\
 MASS\_SUP\_BCG & $\log_{10}(M_\odot)$ & Float & Upper limit of logarithmic stellar mass with 68\% confidence level for the BCG \\
 \tableline
 \end{tabular}
\end{center}
\end{table}

\section{Galaxy clusters with the total mass measurements from X-ray and radio observations} \label{apx:mass} 
Table \ref{tab:mass} lists the galaxy clusters detected by X-ray and SZ-effect observations and their total masses are effectively estimated and rescaled by us to a unified calibration. The characteristic radius of $R_{500}$ is calculated according to Equation (\ref{equ:massr}). 
\begin{table}[!ht]
\begin{center}
\caption{Galaxy clusters with the total mass measurements from X-ray and SZ-effect observations} \label{tab:mass}
 \begin{tabular}{ccccccccc}
 \tableline\tableline
 ID & RA & DEC & Z & M500 & E\_M500 & R500 & E\_R500 &  REF  \\
 (1) & (2) &   (3)  & (4) &   (5)     &      (6)      &   (7)     &     (8)     &   (9)   \\   
 \tableline
   1 &   4.57107 &  16.29432 & 0.55 & 4.11 & 0.77 & 0.93 & 0.06 & W15 \\
   2 &   8.32687 & -21.41319 & 0.19 & 0.99 & 0.20 & 0.66 & 0.04 & T13 \\
   3 &   9.82501 &   0.69981 & 0.28 & 0.66 & 0.13 & 0.56 & 0.04 & W15 \\
   4 &   9.84351 &   0.80273 & 0.41 & 1.62 & 0.31 & 0.72 & 0.05 & T13, T16 \\
   5 &   9.92591 &   0.75922 & 0.42 & 1.27 & 0.25 & 0.66 & 0.04 & T13, T16 \\
   6 &  10.16344 &  25.51840 & 0.15 & 1.24 & 0.24 & 0.72 & 0.05 & T13 \\
   7 &  10.48690 &  25.53105 & 0.13 & 0.93 & 0.19 & 0.66 & 0.04 & T13 \\
   8 &  10.62969 &   0.85368 & 0.16 & 0.69 & 0.15 & 0.59 & 0.04 & T13, T16 \\
   9 &  10.71922 &   0.71671 & 0.27 & 1.16 & 0.23 & 0.68 & 0.04 & T13, T16 \\
  10 &  10.72397 &  -9.57311 & 0.41 & 1.49 & 0.29 & 0.70 & 0.05 & T13 \\
 \tableline
 \tableline
\end{tabular}
\end{center}
\tablecomments{(1) sequence number. (2) right ascension in degree (J2000). (3) declination in degree (J2000). (4) redshift. (5) total mass $M_{500}$ in $10^{14} M_\sun$. (6) error of $M_{500}$ in $10^{14} M_\sun$. (7) $R_{500}$ in Mpc.  (8) error of  $R_{500}$ in Mpc. (9) references: P11 for \citet{pif11}, P15 for \citet{pla16b}, W15 for \citet{wen15}, T13 for \citet{tak13}, T14 for \citet{tak14}, and T16 for \citet{tak16}. \\ 
(This table is available in its entirety in machine-readable form and it can be also available at \url{http://batc.bao.ac.cn/~zouhu/doku.php?id=projects:desi_clusters:start}.) }
\end{table}



\end{document}